\documentclass[fleqn,twoside]{article}
\pdfoutput=1
\usepackage[headings]{espcrc2}
\usepackage{ifpdf}
\usepackage{graphics}
\DeclareGraphicsExtensions{.pdf}
\usepackage{amssymb}
\usepackage{citesort}
\title{Determination of the Jet Energy Scale}
\author{Sven Menke\address[SM]{%
  Max-Planck-Institut f{\"u}r Physik,
  F{\"o}hringer Ring 6, 80805 M{\"u}nchen, Germany}}
\begin{document}
\begin{abstract}
  The uncertainty in jet energy scale is one of the dominating
  systematic errors for many measurements at hadron colliders -- most
  notably for the measurement of the top-quark-mass, inclusive jet
  cross section measurements and last but not least for events with
  large missing transverse energy as expected in searches beyond the
  standard model.  This talk will review the approaches taken at
  Tevatron towards controlling the jet energy scale and discuss
  prospects for the LHC experiments.
\end{abstract}
\maketitle
%

\section{Introduction}
\label{Intro}
Jet energy calibration can roughly be divided in 4 steps:
\begin{description}
\item[Tower and/or cluster reconstruction] Based on the smallest
  calorimeter readout objects (cells or towers) the input for jets are
  formed as fixed size towers or variable size clusters. The latter
  allow for shower containment and shape studies, the former guarantee
  a small size throughout the acceptance. Relevant for the later jet
  calibration is the level of noise- and pile-up-suppression (often
  referred to as zero-suppression) applied to the calorimeter readout
  channels.  Comparisons of events with and without zero-suppression
  are needed to measure the residual background inside jets.
\item[Jet making] The input is grouped in jets defined by proximity in
  $\Delta\eta\times\Delta\phi$ and/or relative transverse momentum.
  Mainly different cone-type~\cite{Blazey:2000qt} and {\tt
    kT}-type~\cite{Ellis:1993tq,Catani:1993hr,Cacciari:2005hq} jet
  algorithms are in use at Tevatron and LHC. Common to all of them is
  the ability to run on any object with $4$-vector interface (cell,
  tower, cluster, track, particle, parton), which allows for comparing
  jets reconstructed on different levels (e.g.~calorimeter jets made
  of towers with MC truth jets made of stable particles). Typical cone
  radii in $\Delta\eta\times\Delta\phi$-space range from $R=0.4$ to
  $R=1.0$ and most modern implementations of cone algorithms address
  infrared and collinear safety. The final jet energy scale is
  jet-algorithm and size dependent but the general calibration
  strategies are independent of the actual algorithm and simply
  repeated for each jet type.
\item[Calibration to stable particle level] Instrumental effects like
  shower containment, noise, and calorimeter response are accounted
  for at this stage. The resulting particle jets can already be used
  for measurements, comparisons with other experiments and theory.
\item[Calibration to parton level] Depending on the analysis and
  physics channel a final correction to the originating parton (quark
  or gluon) is needed. Effects like hadronization, fragmentation,
  flavor-dependency and gluon radiation are taken into account.
\end{description}
\section{Calibration Approaches}
\label{Approaches}
Jets can be either calibrated by applying one or more correction
functions directly on the jet-level (i.e.~depending on the $4$-vector
of the jet) or by calibrating the constituents (cells, towers, or
clusters) and re-calculating the jet energy from the calibrated
constituents. The final calibration from particle to parton level is
applied on the jet $4$-vector in both cases. The current schemes of
the four experiments are:
\begin{description}
\item[CDF] The particle level corrections for the jet in $p_\perp$ are
  given by~\cite{Bhatti:2005ai}:
  \begin{equation}\label{eq:CDF-particle}
    p_\perp^{\textrm{ptcl}} = \left(p_\perp^{\textrm{raw}}
      \times C_\eta-C_{\textrm{MI}}\right) \times C_{\textrm{Abs}},
  \end{equation}
  where $p_\perp^{\textrm{raw}}$ denotes the transverse jet momentum
  on the calorimeter scale, $C_\eta$ corrects for non-uniformities in
  $\eta$ based on simulated di-jet events, $C_{\textrm{MI}}$ removes
  offset energy inside the jet cone due to pile-up, $C_{\textrm{Abs}}$
  scales the jet to the particle level based on di-jet simulations
  with mapped response from matched truth particle jet and calorimeter
  jet pairs, and $p_\perp^{\textrm{ptcl}}$ is the corrected transverse
  jet momentum on particle level.  The corrections are mainly
  simulation driven after tuning the simulation to accurately describe
  the data and uncertainties are derived from validation samples in
  various physics channels (di-jet, $\gamma+\textrm{jet}$, single
  particles).
\item[D0] The energy of the particle jet is given
  by~\cite{Abbott:1998xw,Kvita:2006cm}:
  \begin{equation}\label{eq:D0-particle}
    E^{\textrm{ptcl}} = \frac{\displaystyle E^{\textrm{raw}}-O}{ 
    \displaystyle F_\eta \times R \times S},
  \end{equation}
  where $E^{\textrm{raw}}$ is the jet energy on the calorimeter scale,
  $O$ accounts for energy offsets due to noise and pile-up, $F_\eta$
  corrects for non-uniformities in $\eta$ based on di-jet events, $R$
  scales the jet to the particle level based on $\gamma+\textrm{jet}$
  events, $S$ corrects for the shower containment, and
  $E^{\textrm{ptcl}}$ is the corrected jet energy on particle level.
  The corrections are mainly in-situ driven and separately derived for
  data and simulation.
\item[CMS] A modular scheme similar to that of the Tevatron
  experiments is currently favored~\cite{web:CMS-JES-workshop}, while
  the starting point at initial data taking will be Monte Carlo based
  calibration functions~\cite{DellaNegra:922757}. Another scheme under
  study uses separately clustered but matched deposits in the
  electromagnetic and hadronic sections which are classified as
  "em-like", "hadron-like", or "mip-like" based on the ratio
  $E_{\textrm{had}}/E_{\textrm{em}}$ and calibrated according to
  single particle response~\cite{Bhatti:2006}.
\item[ATLAS] The energy calibration to particle level is currently
  done in two different schemes~\cite{web:ATLAS-HadCalib-workshop}. In
  the first via global calibration functions~\cite{ATLAS:TDRVol1}
  derived from $\chi^2$-fits of di-jet simulations with matched
  calorimeter and particle jets, where weights are applied to the
  individual cell level depending on the energy
  density~\cite{Braunschweig:1987vt} or to the sum of energy deposits
  in each sampling. The second, local hadron calibration method,
  classifies 3d topological clusters~\cite{Cojocaru:2004jk} as
  "em-like" or "hadron-like" by means of cluster shape variables and
  applies H1-style weights~\cite{Issever:2004qh} derived from single
  pion simulations to the cells to account for non-compensation, and
  out-of-cluster weights on the cluster level to account for energy
  deposits in non-clustered cells. In both methods additional
  corrections for energy losses in front or in between samplings are
  applied, again either on global jet or local cluster level.
\end{description}
\begin{figure}[htb]
\centerline{\resizebox{0.4\textwidth}{!}{\includegraphics{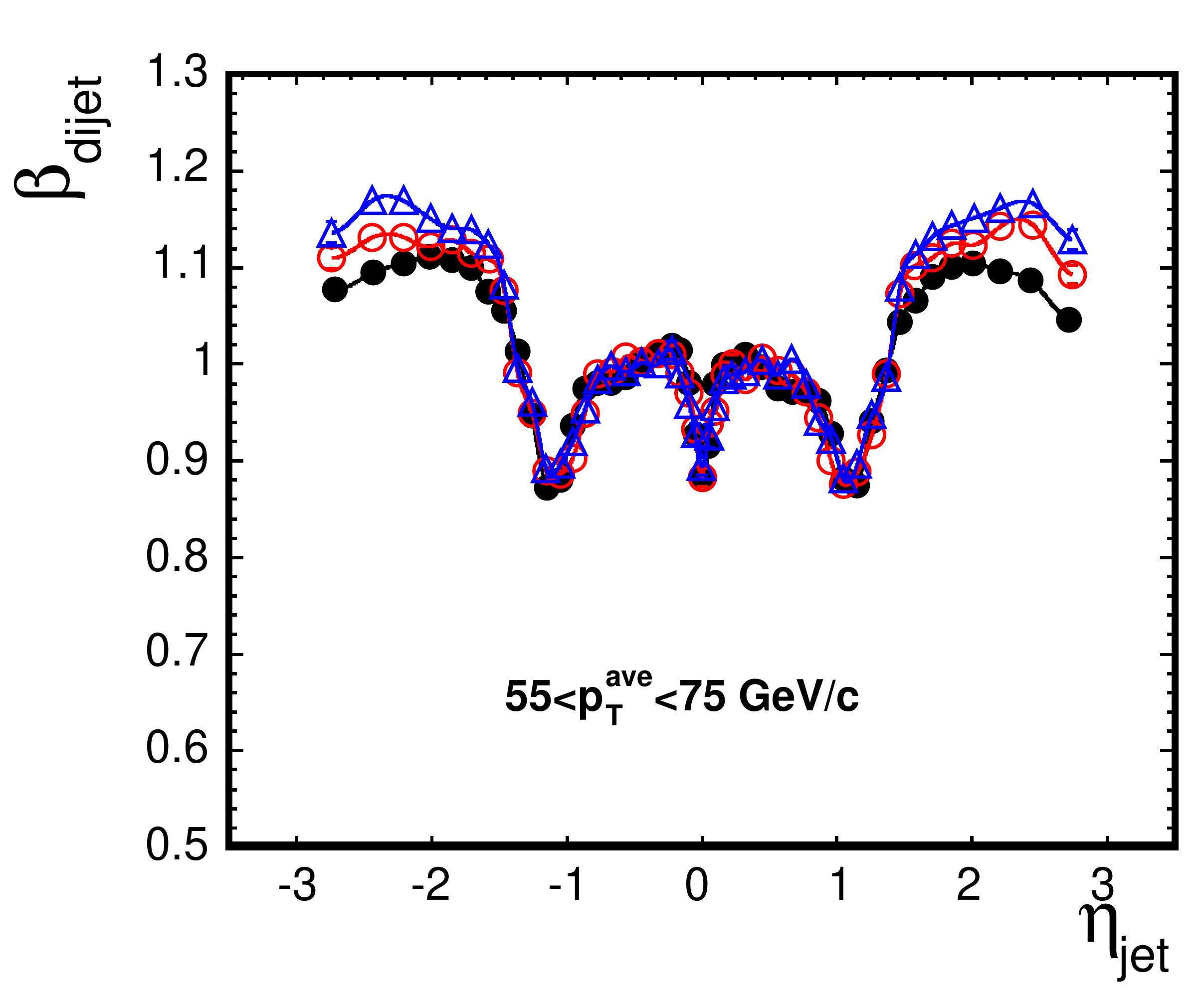}}}
\caption{Di-jet balance for cone jets ($R=0.7$) with
  $55\,{\rm GeV} < \bar{p_\perp} < 75\,{\rm GeV}$ as function of the
  probe jet pseudo-rapidity $\eta_{\textrm{jet}}$ for
  CDF Run II~\cite{Bhatti:2005ai}; solid points show data, open
  triangles Herwig, and open circles Pythia.}
\label{fig:cdfbeta}
\end{figure}
\begin{figure}[htb]
\centerline{\resizebox{0.4\textwidth}{!}{\includegraphics{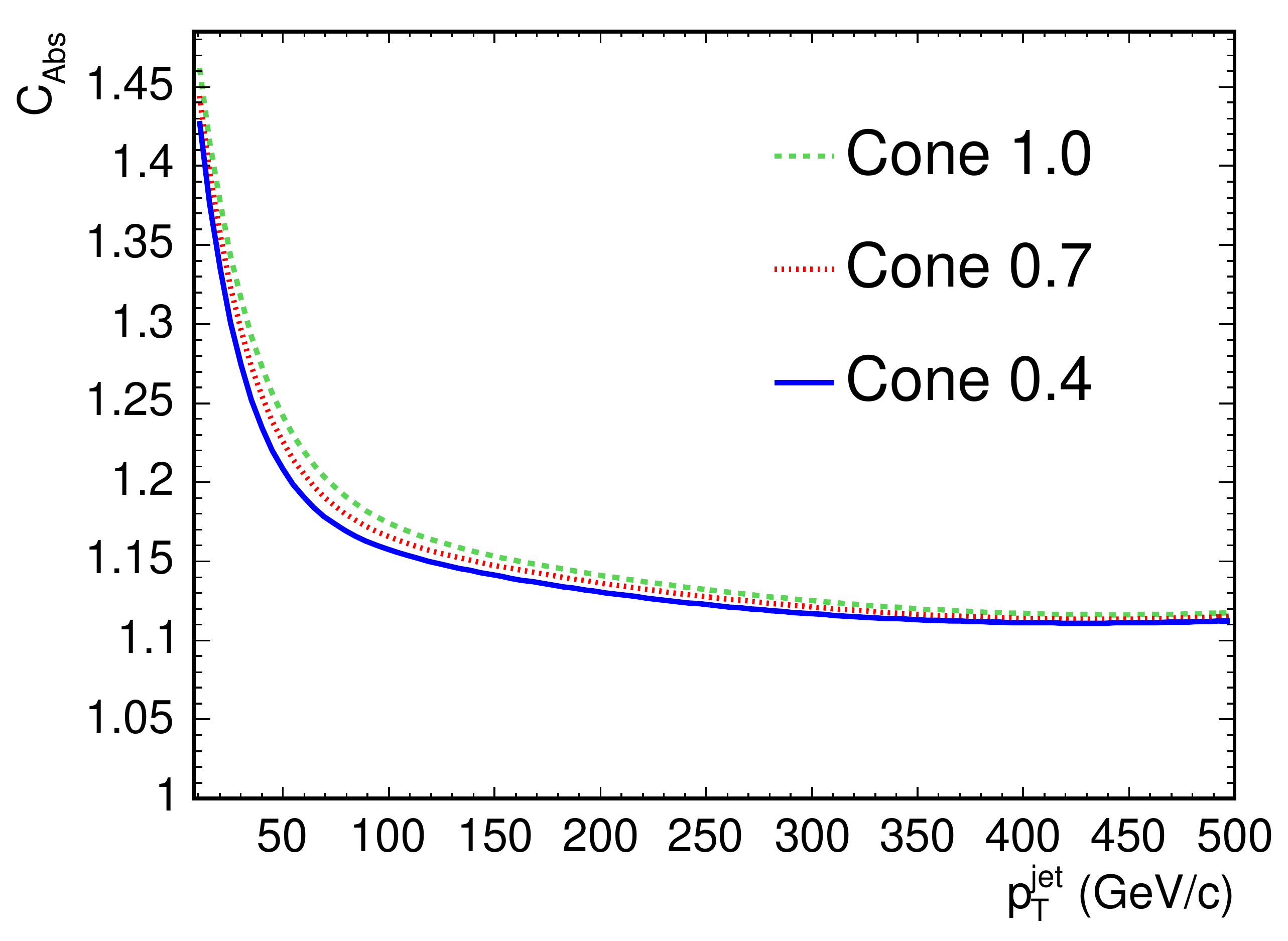}}}
\caption{Absolute jet energy corrections for different cone sizes as
  function of the uncorrected jet transverse momentum
  $p_\perp^{\textrm{jet}}$ for CDF Run II~\cite{Bhatti:2005ai}.}
\label{fig:cdfabs}
\end{figure}
For the simulation driven calibrations the first step is to understand
the calorimeter response to single particles and to tune the Monte
Carlo accordingly. Either Geant4 and/or GFlash based simulations are
in use and compared to beam test data and/or isolated single particles
from minimum bias events.

The most widely used sample to correct for non-uniformities in $\eta$
is QCD $2\to2$ processes where back-to-back requirements in $\phi$ for
the two leading jets and strict upper bounds on the energy of the
sub-leading jets ensure that $p_\perp$-balance arguments hold.  CDF
uses the $p_\perp$ balancing fraction $f_{\textrm{b}} = \Delta
p_\perp/\overline{p_\perp}$ of a central trigger jet and a probe jet
to determine the correction function $C_\eta = (2+\langle
f_{\textrm{b}}\rangle)/(2-\langle f_{\textrm{b}}\rangle)$, which is
shown in Figure~\ref{fig:cdfbeta}.

D0 uses the Missing $E_\perp$ Projection Fraction method (MPF) on
similarly selected di-jet events to obtain the inverse
$\eta$-correction $F_\eta = 1 + \left(\vec{\not{\!\!E}}_\perp \cdot
  \vec{n}_{\textrm{central jet}}\right)/E_{\perp\textrm{\ central
    jet}}$.

For the absolute scale di-jet simulations with matched particle and
relatively corrected calorimeter jets are used in CDF. In D0 the MPF
method is used again but this time on $\gamma+\textrm{jet}$ events,
where the fully reconstructed $\gamma$ drives the absolute scale.
Since only $E_\gamma$ and $\not{\!E}_\perp$ enter the MPF method the
scale is independent of the jet-size but requires additional shower
containment corrections for jets. Figure~\ref{fig:cdfabs} shows the
magnitude 
of the absolute energy scale corrections ranging from about
$10\%$ at high $p_\perp$ to about $50\%$ at low $p_\perp$.
\begin{figure}[h]
  \centerline{\resizebox{0.4\textwidth}{!}{%
    \includegraphics{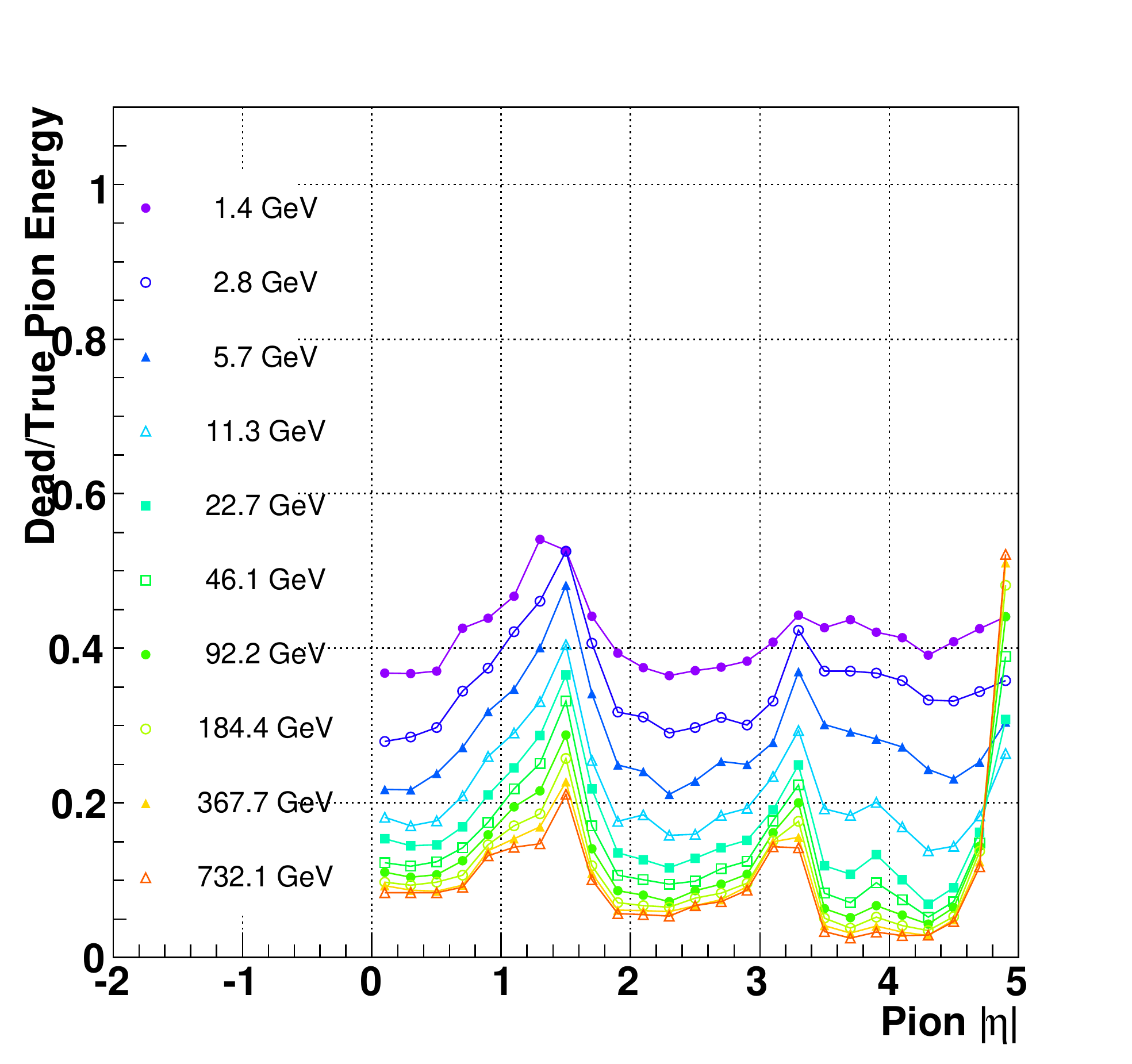}}}
\centerline{\resizebox{0.4\textwidth}{!}{%
    \includegraphics{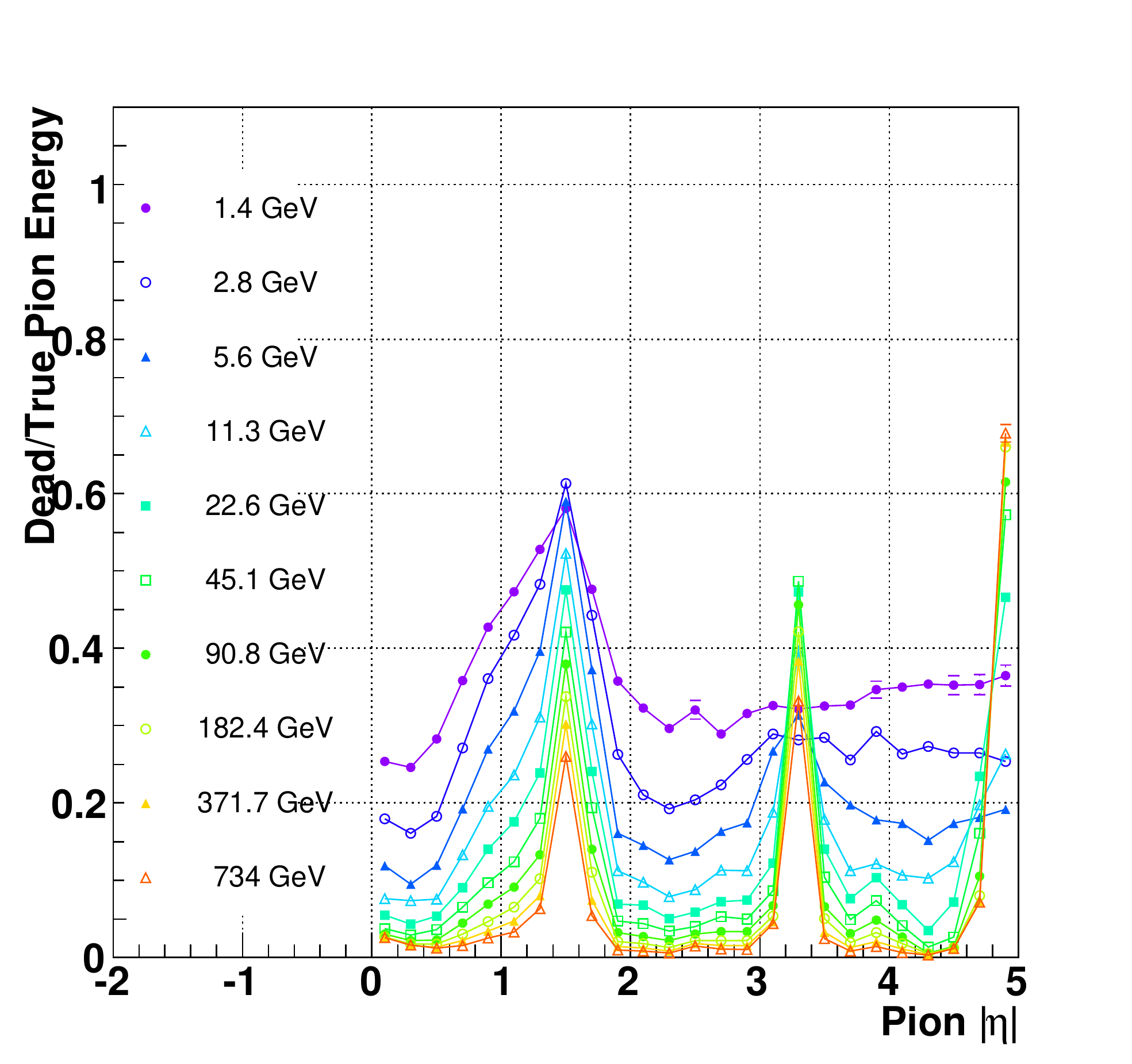}}}
\caption{Energy deposits in dead material in ATLAS from single pion
  simulations at various energies vs. $|\eta|$. The upper plot shows
  charged pions, the lower plot neutral pions.}
\label{fig:deadmat}
\end{figure}
Both relative and absolute scale are taken care of at once in the cell
or sampling level global weighting approaches in ATLAS. The cell
weights are functions of cell energy density and derived for each
sampling separately, where the jet energy dependency enters either as
another argument or is taken care of in a separate jet-level
correction.  The cell weights in the local hadron calibration approach
take a similar form but depend on the cell energy density and cluster
energy instead of the jet energy and are derived from the ratio of
total deposited energy in active and in-active material over
reconstructed energy from single pion simulations. The major
difference here is the treatment of individual small clusters as
particle-like constituents after classifying them as "hadronic" or
"electromagnetic" which allows to apply different type of weights to
different parts of the jet.

The ATLAS approaches require additional corrections for the deposits
in dead material which take the form of sampling energy dependent
weights on the surrounding samplings and are applied on jet or cluster
level. The local hadron calibration allows again to separately correct
"electromagnetic" and "hadronic" clusters, to account for the
different amount of dead-material deposits from neutral and charged
pions as is shown in Figure~\ref{fig:deadmat}.

%
%
If not included already in the applied corrections any cut-offs on the
shower containment due to the finite jet-cone size need to be
corrected for.  D0 separates this effect from physics out-of-cone
effects such as particles radiated outside the jet-acceptance by
looking at the energy deposited in the calorimeter just outside the
calorimeter-jet stemming from particles inside the corresponding
particle-level jet. Similarly the ATLAS local hadron calibration
corrects for finite shower containment on the cluster level taking the
isolation degree of the cluster into account.

%
%
\begin{figure}[htb]
\resizebox{0.44\textwidth}{!}{\includegraphics{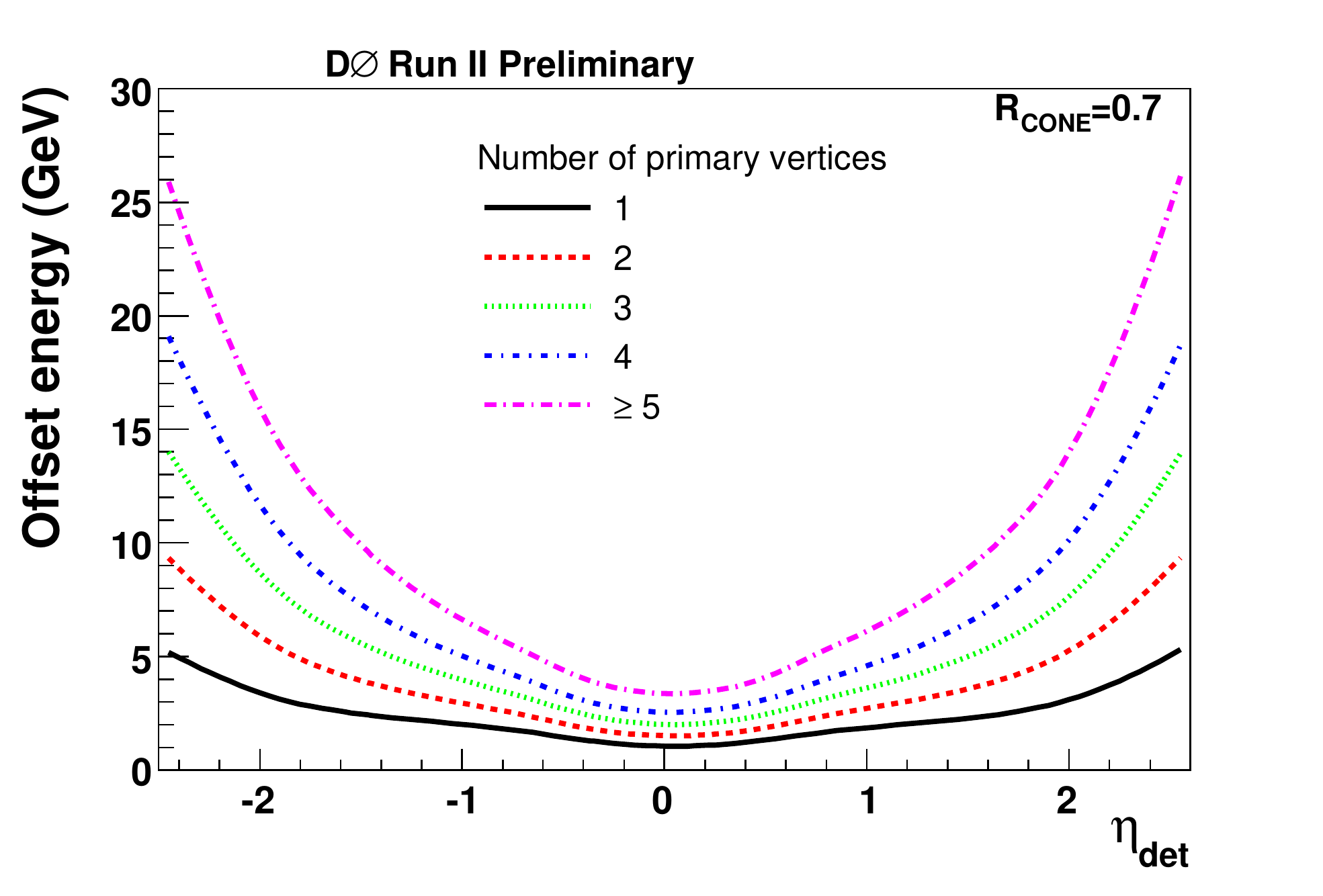}}
\caption{Offset energy for cone jets ($R=0.7$) as function of the jet
  pseudo-rapidity $\eta_{\textrm{det}}$ for different numbers of
  primary vertices in D0 for Run II~\cite{Kvita:2006cm}.}
\label{fig:d0offset}
\end{figure}
Residual noise, underlying event and pile-up contributions to the jets
are subtracted from the jet energy in CDF and D0. The underlying event
and residual noise activity can be measured in low luminosity minimum
bias events using non-zero-suppressed random cones or
$\gamma+\textrm{jet}$ or di-jet events with cones in $\phi$-regions
transverse to the main $\phi$-axis of the hard scatter. A measure of
the pile-up contribution can be parameterized as a function of the
number of observed vertices in minimum-bias events like is shown in
Figure~\ref{fig:d0offset}.

%
%
\begin{figure}[htb]
\resizebox{0.4\textwidth}{!}{\includegraphics{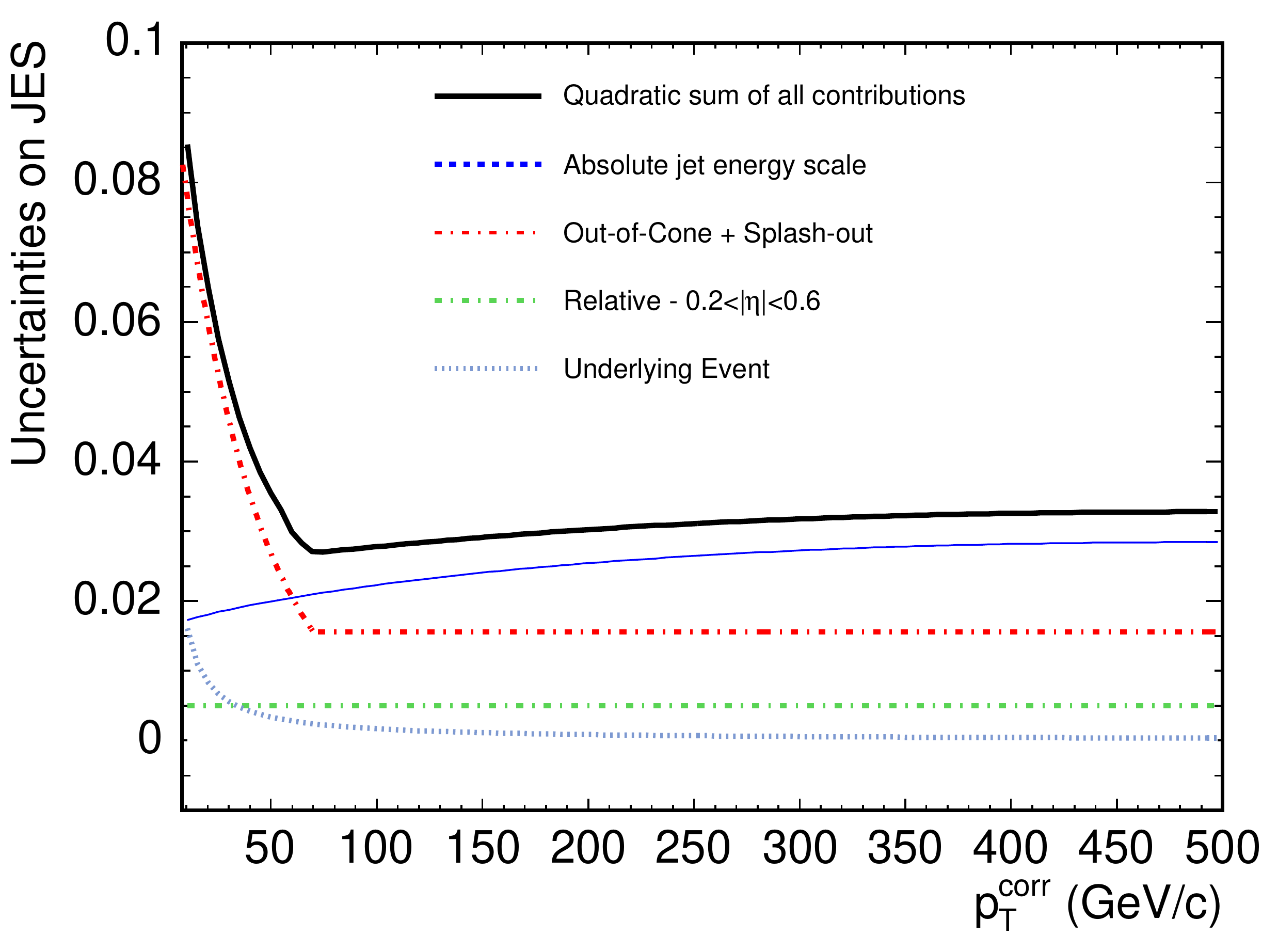}}
\caption{Total Uncertainties on the jet energy scale and the main
  contributions to it for central cone jets ($R=0.4$) as function of
  the corrected transverse jet momentum $p_\perp^{\textrm{corr}}$ in
  CDF for Run II~\cite{Bhatti:2005ai}.}
\label{fig:cdftotsys}
\end{figure}
After the correction to particle level many analyses require to go
back to the parton-level. CDF corrects for physics out-of-cone effects
and underlying event offsets at this stage. The correction is derived
from di-jet simulations as the ratio of the transverse momenta of a
hard parton and the matched particle jet. A crucial step in this
procedure is the validation in $\gamma+\textrm{jet}$-events with
$p_\perp$-balance between parton-level corrected jet and photon
energy. Data driven approaches use the $p_\perp$-balance in
$\textrm{Z}/\gamma+\textrm{jet}$ events directly to obtain the
correction function to parton level.  Also resonant decays like
$\textrm{W}\to \textrm{q}_i\bar{\textrm{q}}_j$ are used to establish
the parton-level scale in-situ by virtue of a mass constraint or via
template methods.
For $\textrm{b}$-jets which have approximately a $5\%$ lower response
in the calorimeters additional corrections from direct comparisons to
the parton level in simulated events are used, but also the
$p_\perp$-balance in $\textrm{Z}/\gamma+\textrm{b-jet}$ and
$\textrm{Z}+\textrm{b-jet}\bar{\textrm{b}}\textrm{-jet}$ events with
leptonically decaying $\textrm{Z}$ are under study.

\section{Uncertainties}
\label{Uncertainties}
The understanding of the jet energy scale reached a very mature level
at Tevatron and uncertainties on $p_\perp$ reach down to about
$3\%$ in the central region as is shown in Figure~\ref{fig:cdftotsys}.
The dominant contributions at the high $p_\perp$ end are from the
absolute energy scale which is still limited by statistics -- either
single hadrons to validate the calorimeter simulation (CDF) or number
of $\gamma+\textrm{jet}$ events (D0). At low $p_\perp$ the
fluctuations in the out-of-cone or shower containment corrections
contribute the largest uncertainty.  Sizeable at low $p_\perp$ are
also the uncertainties introduced from the offset corrections for
noise, pile-up and the underlying event, mainly due to residual
luminosity dependency in each number-of-vertices bin.  This will be
even more substantial at LHC with expected number of multiple
interactions $\sim 10$-times larger at design luminosity.  The aim at
the LHC experiments is to reach a $1\%$ uncertainty level, but a more
realistic number for the start is $4-5\%$.

\section{Conclusions}
\label{Conclusions}
A general trend among all experiments is the modularization of
different calibrations, each dealing with a different effect. The
actual implementations are very different though -- jet based or
constituent based -- Monte Carlo driven or relying on in-situ methods.
For the LHC many options will be pursued in parallel and only the
validation with real data will tell us which ones are to be kept.

\appendix
\section{Acknowledgments}
\label{Acknowledgments}
I'd like to thank the Jet Reconstruction and Jet Energy Scale groups
of ATLAS, CDF, CMS, and D0 for providing me with the material
presented here. In particular I benefited greatly from discussions
with M.~Bosmann, A.~Juste, A.~Kupco, P.~Loch, M.~D'Onofrio, C.~Roda,
N.~Varelas.

\bibliographystyle{elsart-num} \bibliography{hcpproc2007} 
\end{document}